\documentclass[twocolumn,prd,showpacs,aps]{revtex4}

\usepackage{mathrsfs}
\usepackage{amssymb}
\usepackage{latexsym}
\usepackage{hyperref}
\usepackage{amsmath}
\usepackage{graphicx}

\begin{document}
\title{Relativistic Theory of Infinite Statistics Fields}
\author{Chao Cao}\email{ccldyq@gmail.com}
\author{Yi-Xin Chen}\email{yxchen@zimp.zju.edu.cn}
\author{Jian-Long Li}\email{marryrene@gmail.com}
\affiliation{Zhejiang Institute of Modern Physics, Zhejiang
University, Hangzhou 310027, China}

\begin{abstract}
Infinite statistics in which all representations of the symmetric
group can occur is known as a special case of quon theory. However,
the validity of relativistic quon theories is still in doubt. In
this paper we prove that there exists a relativistic quantum field
theory which allows interactions involving infinite statistics
particles. We also give some consistency analysis of this theory
such as conservation of statistics and Feynman rules.

\end{abstract}
 \maketitle

\section{Introduction}
In conventional quantum theory the identical particles always obey
Bose-Einstein statistics or Fermi-Dirac statistics, which are
characterized by commutation or anti-commutation relations
respectively. This restriction in fact requires a symmetrization
postulate that all particles should be in a symmetric state or an
anti-symmetric state \cite{Econdition}. Without such postulate, new
approaches to particle statistics with small violations of Bose or
Fermi statistics are allowed. One famous approach is called quon
theory \cite{quon} in which the annihilation and creation operators
obey the $q$-deformed commutation relation, $a_k a_l^{\dag}
-qa_l^{\dag} a_k=\delta_{kl}, -1\leq q \leq 1$. There exist three
special cases in such theory, Bose and Fermi statistics for $q=\pm
1$, infinite statistics \cite{infinite} for $q=0$.

Infinite statistics with $a_ka^{\dag}_l=\delta_{kl}$ involves no
commutation relation between two annihilation or creation operators.
The quantum states are orthogonal under any permutation of the
identical particles. So it allows all representations of the
symmetric group to occur. Furthermore, the loss of local
commutativity also implies violation of locality, which is an
important character of quantum gravity. By virtue of these
properties, infinite statistics has been applied to many subjects,
such as black hole statistics \cite{BH1, BH2, BH3}, dark energy
quanta \cite{DE1, DE2, DE3, DE4, DE5}, large N matrix theory
\cite{Matrix1, Matrix2, Matrix3} and holography principle \cite{IS,
holography}. Many of these applications involve discussions in
relativistic case.

Unfortunately, the validity of relativistic theory obeying infinite
statistics is still in doubt. Greenberg has showed that the infinite
statistics theory is valid in non-relativistic case. This theory can
also have relativistic kinematics. Cluster decomposition and the CPT
theorem still hold for free fields \cite{quon}. However, there are
two difficulties for infinite statistics to have a consistent
relativistic theory.  For one thing, the physical observables do not
commute at spacelike separation. It is bad news for a relativistic
theory which requires Lorentz invariance for any physical scattering
process (the time-ordering of the operator product in the $S$-matrix
is not Lorentz invariant). For the other thing, by acquiring that
the energies of systems that are widely spacelike separated should
be additive, Greenberg shows that the conservation of statistics in
a relativistic theory limits that $q=\pm 1$, which means it must be
Bose or Fermi case \cite{BF}. The $q=0$ case for infinite statistics
has been excluded.

In this paper we prove the existence of interacting relativistic
field theory obeying infinite statistics by solving the two
difficulties above. First, we directly analyze the Lorentz
invariance of the $S$-matrix from the infinitesimal Lorentz
transformations on $S$. The loss of local commutativity does not
destroy the invariance. Second, we can also show that this field
theory obeys conservation of statistics rule by acquiring some
special form of the interaction Hamiltonian. In addition,  We also
expect that the conventional Feynman rules still hold in this new
theory.

This paper is organized as follows. In Sec. \ref{s2}, we introduce
the elementary ingredients of infinite statistics in
non-relativistic case. In Sec. \ref{s3}, we prove the Lorentz
invariance of the $S$-matrix. In Sec. \ref{s4}, we discuss the
condition that the energies are additive for product states, and
show that the conservation of statistics still holds. Sec. \ref{s5}
discusses the Feynman rules and provides some simple examples.
General conclusions are given in Sec. \ref{s6}.

\section{Infinite Statistics \label{s2}}

The basic algebra of infinite statistics is
\begin{equation} \label{eq:1}
a_ka_l^{\dag}=\delta_{kl},
\end{equation}
where the operator $a_k$ annihilates the vacuum
\begin{equation} \label{eq:2}
a_k|0\rangle=0.
\end{equation}
This relation determines a Fock-state representation in a linear
vector space. m-particle state is constructed as
\begin{equation} \label{eq:3}
|\phi_m\rangle=(a_{k_1})^{m_1}(a_{k_2})^{m_2}...(a_{k_j})^{m_j}|0\rangle,
\end{equation}
with $m_1+m_2+...m_j=m$. Such states have positive norms and the
normalization factor equals one. Since there is no commutation
relation between two annihilation or creation operators, the states
created by any permutations of creation operators are orthogonal.
That's why it is also called ¡°quantum Boltzmann statistics¡±.

One can define a set of number operators $n_i$ such that
\begin{equation} \label{eq:4}
n_i|\phi_m\rangle=m_i|\phi_m\rangle, [n_i, a_j]=-\delta_{ij}a_j.
\end{equation}
Then the total number operator is $N=\sum\limits_{i}n_i$, and the
energy operator is given by $E=\sum\limits_{i}\epsilon_in_i$, where
$\epsilon_i$ is the single particle energy. The explicit form of
$n_i$ was given by Greenberg \cite{infinite}
\begin{equation} \label{eq:5}
\begin{split}
n_i=a_i^{\dag}a_i+\sum\limits_{k}a_k^{\dag}a_i^{\dag}a_ia_k+\sum\limits_{k_1,k_2}a_{k_1}^{\dag}a_{k_2}^{\dag}a_i^{\dag}a_ia_{k_2}a_{k_1}+\cdots\\
+\sum\limits_{k_1,k_2,\ldots,k_s}a_{k_1}^{\dag}a_{k_2}^{\dag}\cdots
a_{k_s}^{\dag}a_i^{\dag}a_ia_{k_s}\cdots a_{k_2}a_{k_1}+\cdots,
\end{split}
\end{equation}
which is obviously a non-local operator. One can easily check that
this definition obeys Eq. ({\ref{eq:4}}).

\section{The Lorentz invariance of the $S$-matrix \label{s3}}

It's not difficult to construct infinite statistics fields that
transform irreducibly under the Lorentz group. In momentum space,
the annihilation field $\psi^+_l(x)$ and creation field
$\psi^-_l(x)$ are
\begin{eqnarray} \label{eq:6}
\psi^{+(n)}_l(x)=\sum\limits_{\sigma n}(2\pi)^{-3/2}\int d^3p\
u_l^{(n)}({\bf
p},\sigma)e^{ip\cdot x}a_p^{(n)}(\sigma),\\
\psi^{-(n)}_l(x)=\sum\limits_{\sigma n}(2\pi)^{-3/2}\int d^3p\
v_l^{(n)}({\bf p},\sigma)e^{-ip\cdot x}a^{\dag(n)}_p(\sigma),
\end{eqnarray}
where $p^\mu$ denotes four-momentum, $\sigma$ labels spin
z-components (or helicity for massless particles), and the
superscript $(n)$ labels particle species
$a_i^{(n)}a_j^{\dag(m)}=\delta(nm)\delta(ij)$. With these fields we
will be able to construct the interaction density as \footnote{See
Chapter 5.1 for details. The derivations of fields such as
$\partial_\mu\psi^{\dag} $ can be seen as new fields with
redefinitions of $u_l,v_l$.}
\begin{equation}\label{eq:8}
\begin{split}
\mathscr{H}(x)=\sum\limits_{N,M}\sum\limits_{n'_1\cdots
n'_N}\sum\limits_{n_1\cdots n_M}\sum\limits_{l'_1\cdots
l'_N}\sum\limits_{l_1\cdots l_M} g_{l'_1\cdots l'_N,l_1\cdots l_M}^{(n'_1\cdots n'_N,n_1\cdots n_M)}\\
\times \psi^{-(n'_1)}_{l'_1}(x)\cdots
\psi^{-(n'_N)}_{l'_N}(x)\psi^{+(n_1)}_{l_1}(x)\cdots
\psi^{+(n_M)}_{l_M}(x).
\end{split}
\end{equation}

In conventional field theory, we usually construct $\mathscr{H}(x)$
out of a linear combination
$\psi(x)=\kappa\psi^+(x)+\lambda\psi^{-c}(x)$, where $c$ denotes
antiparticle. By the requirement of relativistic micro-causality
($[\psi(x), \psi^{\dag}(y)]_{\mp} =0$ for $x-y$ spacelike), we
always have $\kappa=\lambda$. However, in a theory based on infinite
statistics this local commutativity does not hold. {\it So we can't
determine the relationship between $\kappa$ and $\lambda$, and the
basic field in this theory should be $\psi^+(x)$ and $\psi^-(x)$.}
Moreover, from Eq. (\ref{eq:5}) and other operator definitions such
as charge operators one may guess that a general operator
formulation is defined as \cite{infinite}
\begin{equation}\label{eq:9}
\mathcal {A}(\mathcal
{O})=\sum\limits_{m=0}^{\infty}\sum\limits_{n_1,...,n_m,}\sum\limits_{k_1,...,k_m}a^{\dag(n_1)}_{k_1}\cdots
a^{\dag(n_m)}_{k_m}\mathcal{O}\ a^{(n_m)}_{k_m}\cdots
a^{(n_1)}_{k_1}.
\end{equation}
We will see that this definition is important in the next section's
discussion.

With above operator definitions, we can see the Lorentz invariance
of the $S$-matrix. One traditional condition comes from the Dyson
series for the $S$-operator
\begin{equation}\label{eq:10}
\begin{aligned}
S&=T\{{\rm exp}(-i\int_{-\infty}^{\infty}dtV(t))\}\\
 &=1+\sum\limits_{N=1}^{\infty}\frac{(-i)^N}{N!}\int d^4x_1\cdots
 d^4x_N T\{\mathscr{H}(x_1)\cdots \mathscr{H}(x_N)\},
\end{aligned}
\end{equation}
where $V(t)$ is the interaction term $H=H_0+V$ and $T\{\ \}$ denotes
the time-ordered product. Since the time-ordering of two spacetime
points $x_1,x_2$ is invariant unless $x_1-x_2$ is spacelike, so this
condition that makes $S$ Lorentz invariant is that the
$\mathscr{H}(x)$ all commute at spacelike separations
\begin{equation}\label{eq:11}
[\mathscr{H}(x),\mathscr{H}(x')]=0\ \ {\rm for}\ \ (x-x')^2\geq 0.
\end{equation}

Now we compute $[\mathscr{H}(x),\mathscr{H}(x')]$ under infinite statistics
\footnote{Here we consider only one single species of particle which
has no distinct antiparticle for simple. Since other cases can be
treated as involving some intrinsic indices (i.e. adding some delta
functions such as $\delta(mn)$), they won't affect our results.}.
First we write the interaction Hamiltonian density as a polynomial
$\mathscr{H}(x)=\sum\limits_{i}g_{\alpha}\mathscr{H}_{\alpha}(x)$,
each term $\mathscr{H}_{\alpha}$ is a product of definite numbers of
annihilation fields and creation fields. Then we have
\begin{equation}\label{eq:12}
\begin{aligned}
\left[\mathscr{H}(x),\mathscr{H}(x')\right]=&\sum\limits_{\alpha}g^2_{\alpha}[\mathscr{H}_{\alpha}(x),\mathscr{H}_{\alpha}(x')]+\sum\limits_{\alpha
<
\beta}g_{\alpha}g_{\beta}\\&([\mathscr{H}_{\alpha}(x),\mathscr{H}_{\beta}(x')]-[\mathscr{H}_{\alpha}(x'),\mathscr{H}_{\beta}(x)]).
\end{aligned}
\end{equation}
By using Eqs. (\ref{eq:6} - \ref{eq:9}), we have
\begin{equation}\label{eq:13}
\begin{aligned}
&\left[\mathscr{H}_{\alpha}(x),\mathscr{H}_{\alpha}(x')\right]
\\\sim&\int d^3p_1\cdots d^3p_jd^3p'_1\cdots d^3p'_j[\cdots]
\\&a^{\dag}_{p_1}\cdots a^{\dag}_{p_i}(a_{p_{i+1}}\cdots a_{p_j}a^{\dag}_{p'_1}\cdots a^{\dag}_{p'_i})a_{p'_{i+1}}\cdots a_{p'_j}
\\&(e^{i[-(p_1+\cdots+p_i)+(p_{i+1}+p_j)]x+i[-(p'_1+\cdots+p'_i)+(p'_{i+1}+p'_j)]x'}
\\&-(x\leftrightarrow x')),
\end{aligned}
\end{equation}
\begin{equation}\label{eq:14}
\begin{aligned}
&\left[\mathscr{H}_{\alpha}(x),\mathscr{H}_{\beta}(x')]-[\mathscr{H}_{\alpha}(x'),\mathscr{H}_{\beta}(x)\right]
\\\sim&\int d^3p_1\cdots d^3p_jd^3p'_1\cdots d^3p'_l[\cdots]
\\&a^{\dag}_{p_1}\cdots a^{\dag}_{p_i}(a_{p_{i+1}}\cdots a_{p_j}a^{\dag}_{p'_1}\cdots a^{\dag}_{p'_k})a_{p'_{k+1}}\cdots a_{p'_l}
\\&(e^{i[-(p_1+\cdots+p_i)+(p_{i+1}+p_j)]x+i[-(p'_1+\cdots+p'_k)+(p'_{k+1}+p'_l)]x'}
\\&-(x\leftrightarrow x'))+\int d^3p_1\cdots d^3p_ld^3p'_1\cdots d^3p'_j[\cdots]
\\&a^{\dag}_{p_1}\cdots a^{\dag}_{p_k}(a_{p_{k+1}}\cdots a_{p_l}a^{\dag}_{p'_1}\cdots a^{\dag}_{p'_i})a_{p'_{i+1}}\cdots a_{p'_j}
\\&(e^{i[-(p_1+\cdots+p_k)+(p_{k+1}+p_l)]x+i[-(p'_1+\cdots+p'_i)+(p'_{i+1}+p'_j)]x'}
\\-&(x\leftrightarrow x')),
\end{aligned}
\end{equation}
where $[\cdots]$ denotes the product of $u,v,\pi$ factors, $j$, $l$
denote the total numbers of fields in
$\mathscr{H}_{\alpha},\mathscr{H}_{\beta}$, while $i$, $k$ denote
the numbers of creation fields. Since the elements of the $S$-matrix
are the matrix elements of the $S$-operator between free-particle
states
\begin{equation}\label{eq:15}
S_{p'_1p'_2\cdots,p_1p_2\cdots}=\langle0|\cdots
a_{p'_2}a_{p'_1}(S)a^{\dag}_{p_1}a^{\dag}_{p_2}\cdots|0\rangle,
\end{equation}
the condition Eq. (\ref{eq:11}) becomes
\begin{equation}\label{eq:16}
\begin{aligned}
0=&\langle\beta| \int d^4x_1\cdots d^4x_{i-1}d^4x_{i+2}\cdots d^4x_n
 \\&T\{\mathscr{H}(x_1)\cdots[\mathscr{H}(x_i),\mathscr{H}(x_{i+1})]\cdots
 \mathscr{H}(x_n)\}|\alpha\rangle,
 \end{aligned}
\end{equation}
for $(x_i-x_{i+1})^2\geq 0$, where $\langle\beta|,|\alpha\rangle$
denote the final state and the initial state. Those annihilation and
creation operators in  Eqs. (\ref{eq:13}), (\ref{eq:14}) should
contract with $a$'s and $a^{\dag}$'s in the initial states, final
states and other $\mathscr{H}$s (except $\mathscr{H}(x_i)$ and
$\mathscr{H}(x_{i+1})$) in Eq. (16), or they will directly
annihilate the vacuum state and get zero. One should note that the
$S$-matrix involves a four-momentum conservation relation
$S_{\beta\alpha}\sim \delta^4(p_{\beta}-p_{\alpha})$ (see chapter 3
in \cite{weinberg} for details). So after those annihilation and
creation operators in Eqs. (\ref{eq:13}), (\ref{eq:14}) are totally
contracted, we find the commutation relation
$[\mathscr{H}(x),\mathscr{H}(x')]$ is constituted by terms of the
form
\begin{equation}\label{eq:17}
\sim \int\prod d^3k f(k)[e^{i(\sum p+\sum k)(x-x')}-e^{-i(\sum
p+\sum k)(x-x')}],
\end{equation}
where the terms including $k$ come from self-contractions
(contractions don't involve the initial states or final states, such
as contractions in $a_{p_{i+1}}\cdots a_{p_j}a^{\dag}_{p'_1}\cdots
a^{\dag}_{p'_i}$), while $\sum p$ is a sum of some particle momenta
in the initial or final states, which is ,more explicitly, the sum
of momentum in Eqs. (\ref{eq:13}), (\ref{eq:14})
($-(p_1+\cdots+p_i)+(p_{i+1}+p_j)$ or
$-(p_1+\cdots+p_k)+(p_{k+1}+p_l)$) minus the self-contracted momenta
. One can easily see that Eq. (\ref{eq:17}) is non-zero. As a
result, {\it the interaction density $\mathscr{H}(x)$ will not
commute with $\mathscr{H}(x')$ at spacelike separations $x-x'$,
which means that this theory cannot be local}.

However, the failure of the above commutation in $T\{ \}$ of Eq.
(\ref{eq:16}) does not prohibit the existence of a relativistic
field theory. There exists a less restrictive sufficient condition
for Lorentz invariance of the $S$-matrix, which directly comes from
the infinitesimal Lorentz transformations of $S$-operator (see
Chapter 3, page 145 in \cite{weinberg}). This condition is
\begin{equation}\label{eq:18}
0=\int d^3x\int d^3y\  {\bf x}\  [\mathscr{H}({\bf
x},0),\mathscr{H}({\bf y},0)].
\end{equation}
We can also put this condition together with the initial states and
the final states \footnote{More explicitly, the correct formulation
should be $\langle\beta|(\int d^3x\int d^3y\  {\bf x}\
[\mathscr{H}({\bf x},0),\mathscr{H}({\bf
y},0)])f(H,H_0)|\alpha\rangle$, where $f(H,H_0)$ is a polynomial of
$H,H_0$. Since these Hamiltonians won't change the momentum
conservation relation in $[\mathscr{H}({\bf x},0),\mathscr{H}({\bf
y},0)]$ while acting on the initial or final state, we conclude that this
won't affect our results.}. According to our analysis presented
above, this condition becomes
\begin{equation}\label{eq:19}
\begin{aligned}
0=&\int\prod d^3k f(k)\int d^3x\int d^3y\  {\bf x}\\ &(e^{i(\sum
{\bf p}+\sum {\bf k})({\bf x-y})}-e^{-i(\sum {\bf p}+\sum {\bf
k}){\bf (x-y)}})
\\=&\int\prod d^3k f(k)\delta(\sum {\bf p}+\sum {\bf k})\\&\int d^3x \ {\bf x} \ (e^{i(\sum {\bf p}+\sum {\bf k}){\bf
x}}-e^{-i(\sum {\bf p}+\sum {\bf k}){\bf x}}),
\end{aligned}
\end{equation}
which is always satisfied because $\int d^3x \  {\bf x}e^{i{\bf
p}{\bf x}}$ is even in ${\bf p}$. {\it So we conclude that the
interacting field theory based on infinite statistics is Lorentz
invariant}.

A similar analysis can be applied to the commutation relation
$[\mathcal{A}(\mathscr{H}(x)),\mathcal{A}(\mathscr{H}(x'))]=\mathcal{A}(\mathscr{O}(x,x'))$
(see Appendix for details), in which the interaction density
$\mathcal{A}(\mathscr{H})$ is defined in Eq. (\ref{eq:9}). This
commutation can be decomposed into a sum of terms that are similar
to Eqs. (\ref{eq:13}), (\ref{eq:14}). Noting that the momenta of
$a$'s and $a^{\dag }$'s summed over in these terms (such as $p_i$ in
$\sum a^\dag_{p_i}\cdots\mathscr{O}(x,x')\cdots a_{p_i}$ terms) have
no contribution to the momentum conservation relation in
$\mathscr{O}$, the final terms after total contraction are still as
the form (\ref{eq:17}). Then by using condition (\ref{eq:18}),{\it
we conclude that the theory with interaction density of the form
$\mathcal{A}(\mathscr{H}(x))$ is also Lorentz invariant}.

\section{Conservation of statistics\label{s4}}
Here we try to impose the condition that the energy should be
additive for product states on the interaction $\mathscr{H}(x)$. For
subsystems that are widely spacelike separated, the contribution to
the energy should be additive if \cite{Econdition}
\begin{equation}\label{eq:21}
[\mathscr{H}(x), \psi(x')]\rightarrow 0, \ {\rm as} \
x-x'\rightarrow \infty \ {\rm spacelike}
\end{equation}
for all fields. One can check that this condition is equivalent to
\begin{equation}\label{eq:22}
(e^{iH_At}\psi_A)\otimes
(e^{iH_Bt}\psi_B)=e^{i(H_A+H_B)t}(\psi_A\otimes\psi_B),
\end{equation}
while subsystems $A$ and $B$ are widely spacelike separated. By
using Eq. (\ref{eq:21}), Greenberg expected that the Hamiltonian
operators should be effectively bosonic, which leads to
``conservation of statistics" and acquires that $q=\pm1$ \cite{quon,
BF}. However, we think this restriction is too strong, and we
provides a much less restrictive requirement on $\mathscr{H}(x)$,
which also leads to conservation of statistics.

In order to satisfy the energy additive condition, we should replace
the density $\mathscr{H}$  with $\mathcal{A}(\mathscr{H})$. Thus Eq.
(\ref{eq:21}) becomes
\begin{equation}\label{eq:20}
[\mathcal{A}(\mathscr{H}(x)), \psi(x')]\rightarrow 0, \ {\rm as} \
x-x'\rightarrow \infty \ {\rm spacelike}.
\end{equation}
Noting that the basic fields are $\psi^{\pm}$ here, we can get
\begin{equation}\label{eq:23}
\begin{aligned}
&[\mathcal{A}(\mathscr{H}(x)),\psi^{+}(x')]=-\psi^{+}(x')\mathscr{H}(x)\\
&[\mathcal{A}(\mathscr{H}(x)),\psi^{-}(x')]=\mathscr{H}(x)\psi^{-}(x').
\end{aligned}
\end{equation}
We note that in infinite statistics
\begin{equation}\label{eq:24}
\psi^{+(n)}(x)\psi^{-(m)}(x')\sim[\cdots]\delta(nm)\Delta_+(x-x'),
\end{equation}
where the coefficient $[\cdots]$ may contain some derivative times
such as $\gamma^{\mu}\partial_{\mu}$ for spin $\frac{1}{2}$
particles and $\partial^{\mu}\partial_{\mu}$ for spin one particles,
while
$\Delta_+(x-x')\equiv\frac{1}{(2\pi)^3}\int\frac{d^3p}{2p^0}e^{ip(x-x')}$
is a standard function \footnote{One can see Chapter 5 in
\cite{weinberg} for more details. Since we have proved the existence
of relativistic infinite statistics field theory, those analyzes
based on Lorentz group are still valid.}. Moreover, for
$(x-x')^2\geq0$
\begin{equation}\label{eq:25}
\Delta_+(x-x')=\frac{m}{4\pi^2\sqrt{(x-x')^2}}K_1(m\sqrt{(x-x')^2}),
\end{equation}
in which $K_1()$ is the modified Hankel function of order 1. So
$\Delta_+(x-x')$ and its derivations are $\rightarrow 0$ as
$x-x'\rightarrow \infty$ spacelike. {\it Then by using Eqs.
(\ref{eq:23}) and (\ref{eq:24}), we infer that the condition
(\ref{eq:20}) is satisfied if the interaction density
$\mathscr{H}(x)$ has the form Eq. (\ref{eq:8}) with $N,M\geq1$}.
Moreover, in order to get condition (\ref{eq:22}) satisfied for our
new definition $\mathcal{A}(H)$, we also need
$[\mathcal{A}(\mathscr{H}(x)),
\mathcal{A}(\mathscr{H}(x'))]\rightarrow 0, \ {\rm as} \
x-x'\rightarrow \infty \ {\rm spacelike}$, this proof is given in
Appendix. Here we just exclude the terms in $\mathscr{H}$ which are
products containing only annihilation fields $\psi^+(x)$ (or
creation fields $\psi^-(x)$). Since the operation definition
$\mathcal{A}(\mathscr{H})$ is quite normal in infinite statistics
field theory \cite{infinite}, our requirement is much looser than
the condition that Hamiltonian operator must be effectively bosonic.

Although the interaction density may not be bosonic, conservation of
statistics still holds in our theory. To see this, let's consider
the case that infinite statistics fields couple to normal fields (we
will denote infinite statistics fields by the subscript $I$ and
normal statistics fields by the subscript $B$). According to
conventional fields theory and our above discussion, all
interactions must involve any number of bosons, an even number of
fermions (including zero), at least one annihilation infinite
statistics field and at least one creation infinite statistics
field. These three kinds of particles commute with each other, so
$\mathcal{A}(\mathcal{O}_I\mathcal{O'}_B)=\mathcal{A}(\mathcal{O}_I)\mathcal{O'}_B$.
Since we exclude the terms in $\mathscr{H}$ which are products
containing only annihilation fields (or creation fields), then the
term $T\{\mathcal{A}(\mathscr{H}(x_1))\cdots
\mathcal{A}(\mathscr{H}(x_N))\}$ in $S$-operator (\ref{eq:10}) must
have both $\psi^+_I$ and $\psi^-_I$ fields after
$a_{Ii}^{(n)}a_{Ij}^{\dag(m)}=\delta(nm)\delta(ij)$ contractions. So
there must be infinite statistics particles both in the initial and
final states, which forbids any process that the in-state obeys
infinite statistics (normal statistics) while the out-state obeys
normal statistics (infinite statistics). Moreover, since the
interaction vertices such as
$\mathcal{A}(\psi_I^+\psi_I^{+c})\psi_B$ and
$\mathcal{A}(\psi_I^+)(\psi_B^{\dag}\psi_B)$ do not exist, we also
exclude those virtual processes such as annihilation of a pair of
infinite (normal) statistics particles producing a normal (infinite)
statistics particle, which also break conservation of statistics. So
{\it we conclude that our theory obeys the conservation of
statistics}. Some examples are presented in in the next section.

\section{Feynman rules and Examples\label{s5}}

In order to derive Feynman rules, first we see ``Wick's theorem''
for infinite statistics fields, by using the relation
$a_{Ii}^{(n)}a_{Ij}^{\dag(m)}=\delta(nm)\delta(ij)$, any product of
a set of infinite statistics operators can be finally expressed as a
normal product. This looks a bit different from conventional field
theory, in which contractions can arise between any creator and
annihilator pairs by permuting the operators, while in our theorem
contractions can only arise between the neighboring operators.
However, by inducing $\mathcal{A}(\mathscr{H})$, we can also realize
some permutations. For example, in order to get contraction between
a final particle $\langle 0|\cdots a$ and a creation field
$\psi^{-}(y)$ in $\langle 0|\cdots
a(\mathcal{A}(\psi^-(x)\psi^+(x))\mathcal{A}(\psi^-(y)\psi^+(y))\cdots|0\rangle$,
we can use
$\sum\limits_{i}a_i^{\dag}(\psi^-(x)\psi^+(x))a_i(\psi^-(y)\psi^+(y))=\psi^-(y)(\psi^-(x)\psi^+(x))\psi^+(y)$
to move $\psi^{-}(y)$ to the left. {\it So by using the operator
definition (\ref{eq:9}), we can get ``Wick's theorem'' for infinite
statistics fields.}

Since the operators can't be moved arbitrarily, there will be some
limits on the Feynman rules. In fact, the step functions $\theta(x)$
do not just appear in propagators, but also affect the external
lines. To see this, let's take
$\mathscr{H}=\psi^-\psi^-\psi^++\psi^-\psi^+\psi^+$ for example.
Then $S$-operator contains a term
$\theta(x-y)(\psi^-(x)\psi^-(x)\psi^+(x))(\psi^-(y)\psi^+(y)\psi^+(y))+\theta(y-x)(\psi^-(y)\psi^-(y)\psi^+(y))(\psi^-(x)\psi^+(x)\psi^+(x))$,
in which $\psi^-(x)\psi^-(x)$ and $\psi^+(y)\psi^+(y)$ are
unexchangable. If the final state is $\langle p_1p_2|$ and the
initial state is $|p_3p_4\rangle$, such term will be a sum of two
subgraphs, one with two external lines carrying momenta $p_1,p_2$ at
$x$ point, two external lines carrying momenta $p_3,p_4$ at $y$
point and one internal line $\Delta(x-y)$, while the other with two
external lines carrying momenta $p_3,p_4$ at $x$ point, two external
lines carrying momenta $p_1,p_2$ at $y$ point and one internal line
$\Delta(y-x)$. The propagator $\Delta(x-y)$ is defined as
\begin{equation}\label{eq:26}
\begin{aligned}
&\ \Delta_{lm}(x-y)\\\equiv&\ i\theta(x-y)(\psi^+_l(x)\psi^-_m(y))
\\=&\ (2\pi)^4\int d^4q \frac{-P_{lm}^{(L)}(q)e^{iq(x-y)}}{2\sqrt{{\bf q}^2+m^2}(q^0-\sqrt{{\bf
q}^2+m^2}+i\epsilon)},
\end{aligned}
\end{equation}
in which $P_{lm}^{(L)}$ is defined in Chapter 6.2 in
\cite{weinberg}. Noting that the position related term $e^{iq(x-y)}$
is still the same as in conventional propagator, we can infer that
the Feynman rules for external lines in momentum space (after
integrating over the spacetime position $x,y$) are the same as
before. So the contribution of the external lines to the $S$-matrix
are the same in the two subgraphs. If we denote
$\Delta_q=(2\pi)^4\int d^4q (-P_{lm}^{(L)}(q))/(2\sqrt{{\bf
q}^2+m^2}(q^0-\sqrt{{\bf q}^2+m^2}+i\epsilon))$ as internal line
contribution in momentum space, then the total $S$-matrix for this
$1\ 2\rightarrow 1'\ 2'$ progress is $(external\ line\ terms) \cdot
(\Delta_q+\Delta_{-q})=(external\ line\ terms) \cdot \Delta_F$, in
which $\Delta_F$ is the conventional propagator in momentum space.
{\it So we expect that the Feynman rules for our new theory are
still the same as before.} This allows us to apply some traditional
methods such as renormalization analysis.

Here we give two examples for infinite statistics field
interactions. For simple, we consider a scattering process $1\
2\rightarrow 1'\ 2'$ and the interaction density is trilinear in a
set of real scalar fields. First, for pure infinite statistics
interaction, we take
$\mathscr{H}=\phi^-\phi^-\phi^++\phi^-\phi^+\phi^+$, the Feynman
diagrams are presented in Fig. 1. Fig. 1.(a) describes the term
$\theta(x-y)(\phi^-(x)\phi^-(x)\phi^+(x))(\phi^-(y)\phi^+(y)\phi^+(y))+x\leftrightarrow
y$; Fig. 1.(b) describes the term
$\theta(x-y)\sum\limits_{i}[(\phi^-(x)\phi^+(x)\phi^+(x))a^{\dag}_i(\phi^-(y)\phi^-(y)\phi^+(y))a_i+a^{\dag}_i(\phi^-(x)\phi^+(x)\phi^+(x))a_i(\phi^-(y)\phi^-(y)\phi^+(y))]+x\leftrightarrow
y$. Secondly, for the case that infinite statistics fields couple to
a bosonic field, we take $\mathscr{H}=\phi^-_I\phi^+_I\phi_B$, the
Feynman diagrams are presented in Fig. 2. Fig. 2.(a)(b) describe the
term
$\theta(x-y)(\phi_B(x)\phi^-_I(x)\phi^+_I(x))(\phi^-_I(y)\phi^+_I(y)\phi_B(y))+x\leftrightarrow
y$; Fig. 2.(c) describes the term
$\theta(x-y)[\phi_B^+(x),\phi_B^-(y)]\sum\limits_{i}[a^{\dag}_{Ii}(\phi^-_I(x)\phi^+_I(x))a_{Ii}(\phi^-_I(y)\phi^+_I(y))+(\phi^-_I(x)\phi^+_I(x))a^{\dag}_{Ii}(\phi^-_I(y)\phi^+_I(y))a_{Ii}]+x\leftrightarrow
y$. We see that those processes breaking the conservation of
statistics as presented in Fig. 3 are excluded.

\begin{figure}[t]
\includegraphics[scale=0.8]{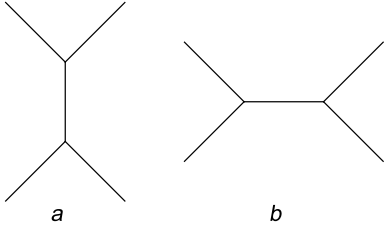}
\caption{Pure infinite statistical interactions} \label{fig:1}
\end{figure}
\begin{figure}[t]
\includegraphics[scale=0.8]{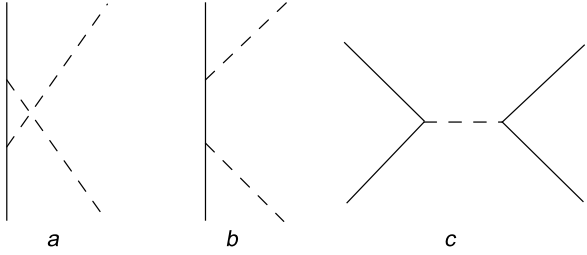}
\caption{Interactions between Bose and infinite statistics. The
solid lines represent infinite statistical particles, and the dashed
lines represent bosons.} \label{fig:2}
\end{figure}
\begin{figure}[t]
\includegraphics[scale=0.8]{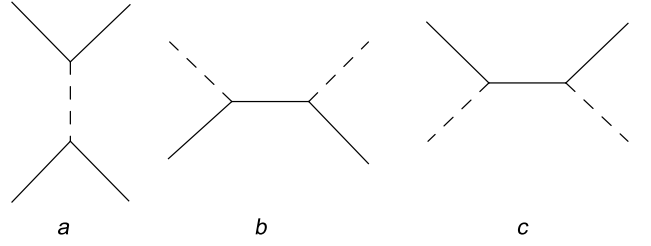}
\caption{Interactions violating the conservation of statistics
rules. The solid lines represent infinite statistical particles, and
the dashed lines represent bosons.} \label{fig:3}
\end{figure}

\section{Conclusions\label{s6}}
In this paper, we have showed that the scattering processes
involving infinite statistics particles are Lorentz invariant. This
proof is directly based on the infinitesimal Lorentz transformations
on the $S$-matrix. By applying the condition that the energies are
additive for product states, we have showed this theory can obey
conservation of statistics with selected interaction Hamiltonian.
For all the above reasons, we conclude that the relativistic quantum
field theory can also involve infinite statistics particles. For
infinite statistics part of this theory, the basic fields should be
both annihilation fields $\psi^+(x)$ and creation fields
$\psi^-(x)$, while the interaction density has a nonlocal definition
$\mathcal{A}(\mathscr{H}(x))$ in which $\mathscr{H}(x)$ should take
the form Eq. (\ref{eq:8}) with $N, M\geq1$. Since we have showed
that the conventional Feynman rules are still valid, some
traditional methods such as renormalization analysis can also be
extended to our new theory.

\acknowledgments

This work is supported in part by the NNSF of China Grant No.
90503009, No. 10775116, and 973 Program Grant No. 2005CB724508.

\appendix*\section{}

 In this Appendix, we calculate the commutation relation
$[\mathcal{A}(\mathscr{H}(x)),\mathcal{A}(\mathscr{H}(x'))$. Here we
also consider one single species of particle for simple. As we have
shown in Sec. \ref{s3}, this commutation can be decomposed into a
sum of
$g_{\alpha}g_{\beta}[\mathcal{A}(\mathscr{H_{\alpha}}(x)),\mathcal{A}(\mathscr{H_{\beta}}(x'))]$,
in which $\mathscr{H_{\alpha}}$, $\mathscr{H_{\beta}}$ are products
of definite numbers of annihilation fields and creation fields. Here
we denote the total numbers of fields in $\mathscr{H_{\alpha}}$,
$\mathscr{H_{\beta}}$ are $j$, $l$, while the numbers of creation
fields are $i$, $k$. Then we only need to calculate the relation $
[\mathcal{A}(a^{\dag}_{p_1}\cdots a^{\dag}_{p_i}a_{p_{i+1}}\cdots
a_{p_j}), \mathcal{A}(a^{\dag}_{p'_1}\cdots
a^{\dag}_{p'_k}a_{p'_{k+1}}\cdots a_{p'_l})]$.

Firstly, we write the product $\mathcal{A}(a^{\dag}_{p_1}\cdots
a^{\dag}_{p_i}a_{p_{i+1}}\cdots a_{p_j})\cdot
\mathcal{A}(a^{\dag}_{p'_1}\cdots a^{\dag}_{p'_k}a_{p'_{k+1}}\cdots
a_{p'_l})$ as
\begin{equation}\label{app:1}
\begin{aligned}
&\mathcal{A}(a^{\dag}_{p_1}\cdots a^{\dag}_{p_i}a_{p_{i+1}}\cdots
a_{p_j})\cdot \mathcal{A}(a^{\dag}_{p'_1}\cdots
a^{\dag}_{p'_k}a_{p'_{k+1}}\cdots a_{p'_l})\\=
&\mathscr{O}_{11}+\mathscr{O}_{12}+\mathscr{O}_{13}+... +\mathscr{O}_{1n}+... \\
& \mathscr{O}_{21}+\mathscr{O}_{22}+\mathscr{O}_{23}+... +\mathscr{O}_{2n}+...\\
& \mathscr{O}_{31}+\mathscr{O}_{32}+\mathscr{O}_{33}+... +\mathscr{O}_{3n}+...\\
& \vdots \\
& \mathscr{O}_{m1}+\mathscr{O}_{m2}+\mathscr{O}_{m3}+... +\mathscr{O}_{mn}+... \\
& \ldots,
\end{aligned}
\end{equation}
in which $\mathscr{O}_{mn}$ is defined as the product of the $m$-th
term in $\mathcal{A}(\mathscr{H_{\alpha}})$ and the $n$-th term in
$\mathcal{A}(\mathscr{H_{\beta}})$
\begin{equation}\label{app:2}
\begin{aligned}
&\ \ \mathscr{O}_{mn}\\\equiv& (\sum\limits_{q}a^{\dag}_{q_1}\cdots
a^{\dag}_{q_{m-1}}(a^{\dag}_{p_1}\cdots
a^{\dag}_{p_i}a_{p_{i+1}}\cdots a_{p_j}) a_{q_{m-1}}\cdots
a_{q_1})\cdot\\&(\sum\limits_{q'}a^{\dag}_{q'_1}\cdots
a^{\dag}_{q'_{n-1}}(a^{\dag}_{p'_1}\cdots
a^{\dag}_{p'_k}a_{p'_{k+1}}\cdots a_{p'_l}) a_{q'_{n-1}}\cdots
a_{q'_1}).
\end{aligned}
\end{equation}
It's not difficult to find that $\mathscr{O}_{mn}$ have a recursion
relation
\begin{equation}\label{app:3}
\mathscr{O}_{(m+1) (n+1)}=\sum_q a^{\dag}_q \mathscr{O}_{mn} a_q.
\end{equation}
Thus we can denote
$\mathscr{O}_{mn}+\mathscr{O}_{(m+1)(n+1)}+\mathscr{O}_{(m+2)(n+2)}+\cdots
+\mathscr{O}_{(m+s)(n+s)}+\cdots$ by
$\mathcal{A}({\mathscr{O}_{mn}})$. Then the product (\ref{app:1})
can be simplified as
\begin{equation}\label{App:4}
\begin{aligned}
&\mathcal{A}(a^{\dag}_{p_1}\cdots a^{\dag}_{p_i}a_{p_{i+1}}\cdots
a_{p_j})\cdot \mathcal{A}(a^{\dag}_{p'_1}\cdots
a^{\dag}_{p'_k}a_{p'_{k+1}}\cdots a_{p'_l}) \\= &
\mathcal{A}({\mathscr{O}_{11}})+\mathcal{A}({\mathscr{O}_{12}})+\mathcal{A}({\mathscr{O}_{13}})+\cdots
+
\mathcal{A}({\mathscr{O}_{1n}})+\cdots\\
&
\mathcal{A}({\mathscr{O}_{21}})+\mathcal{A}({\mathscr{O}_{31}})+\mathcal{A}({\mathscr{O}_{41}})+\cdots
+ \mathcal{A}({\mathscr{O}_{m1}})+\cdots.
\end{aligned}
\end{equation}
We can also define $\mathscr{O'}_{mn}$ as the product of the $m$-th
term in $\mathcal{A}(\mathscr{H_{\beta}})$ and the $n$-th term in
$\mathcal{A}(\mathscr{H_{\alpha}})$. Then the commutation becomes
\begin{equation}\label{App:5}
\begin{aligned}
& [\mathcal{A}(a^{\dag}_{p_1}\cdots a^{\dag}_{p_i}a_{p_{i+1}}\cdots
a_{p_j}), \mathcal{A}(a^{\dag}_{p'_1}\cdots
a^{\dag}_{p'_k}a_{p'_{k+1}}\cdots a_{p'_l})] \\= &
\mathcal{A}({\mathscr{O}_{11}})+\mathcal{A}({\mathscr{O}_{12}})+\mathcal{A}({\mathscr{O}_{13}})+\cdots
+
\mathcal{A}({\mathscr{O}_{1n}})+\cdots \\
&
\mathcal{A}({\mathscr{O}_{21}})+\mathcal{A}({\mathscr{O}_{31})}+\mathcal{A}({\mathscr{O}_{41}})+\cdots
+
\mathcal{A}({\mathscr{O}_{m1}})+\cdots \\
&
-\mathcal{A}({\mathscr{O}^\prime_{11}})-\mathcal{A}({\mathscr{O}^\prime_{12}})-\mathcal{A}({\mathscr{O}^\prime_{13}})-\cdots
-\mathcal{A}({\mathscr{O}^\prime_{1n}})-\cdots \\
&-\mathcal{A}({\mathscr{O}^\prime_{21}})-\mathcal{A}({\mathscr{O}^\prime_{31}})-\mathcal{A}({\mathscr{O}^\prime_{41}})-\cdots
- \mathcal{A}({\mathscr{O}^\prime_{m1}})-\cdots \\
=&\mathcal{A}(({\mathscr{O}_{11}}+\cdots +
{\mathscr{O}_{1(j-i)}}+{\mathscr{O}_ {21}}+\cdots +
{\mathscr{O}_{k1}})\\&-({\mathscr{O}^\prime_{11}}+\cdots
+{\mathscr{O}^\prime_{1(l-k)}}+{\mathscr{O}^\prime_{21}}+\cdots
+{\mathscr{O}^\prime_{i1}})).
\end{aligned}
\end{equation}
All the other terms are canceled by the relation
\begin{equation}\label{app:6}
\begin{aligned}
&\mathcal{A}({\mathscr{O}_{1(j-i+m)}})-\mathcal{A}({\mathscr{O}^\prime_{(i+m)
1}})=0\\&\mathcal{A}({\mathscr{O}_{(k+m)1}})-\mathcal{A}({\mathscr{O}^\prime_{1(l-k+m)}})=0\
\  m\geq1.
\end{aligned}
\end{equation}

$\mathscr{O}_{mn}$ (or $\mathscr{O'}_{mn}$) in the remaining $j+l-2$
terms are some permutations of $(a^{\dag}_{p_1}\cdots
a^{\dag}_{p_i}a_{p_{i+1}}\cdots a_{p_j})\cdot(a^{\dag}_{p'_1}\cdots
a^{\dag}_{p'_k}a_{p'_{k+1}}\cdots a_{p'_l})$. Moreover, if the
interaction density $\mathscr{H}(x)$ is defined in Eq. (\ref{eq:8})
with $N,M\geq1$, then by using Eq. (\ref{App:5}) each term in
$[\mathcal{A}(\mathscr{H}(x)), \mathcal{A}(\mathscr{H}(x'))]$
involves $\psi^+(x)\psi^-(x')$ (or $\psi^+(x')\psi^-(x)$). So
$[\mathcal{A}(\mathscr{H}(x)),
\mathcal{A}(\mathscr{H}(x'))]\rightarrow 0, \ {\rm as} \
x-x'\rightarrow \infty \ {\rm spacelike}$, which is a necessary
condition for Eq. (\ref{eq:22}).

\end{document}